\documentclass[a4paper]{jpconf}
\usepackage{graphicx}
\usepackage{amssymb} 
\usepackage{subcaption} 
\usepackage{hyperref} 
\begin{document}
\title{A high-performance custom photodetection system to probe the light yield enhancement in oriented crystals}

\author{
    M~Soldani$^{1, 2}$,
    L~Bandiera$^2$,
    L~Bomben$^{3, 4}$,
    C~Brizzolari$^{4, 5}$,
    R~Camattari$^{1, 2}$,
    D~De~Salvador$^{6, 7}$,
    V~Guidi$^{1, 2}$,
    V~Haurylavets$^8$,
    E~Lutsenko$^{3, 4}$,
    T~Maiolino$^9$,
    V~Mascagna$^{3, 4}$,
    A~Mazzolari$^2$,
    M~Prest$^{3, 4}$,
    M~Romagnoni$^{2, 10}$,
    F~Ronchetti$^{3, 4}$,
    A~Selmi$^{3, 4}$,
    A~Sytov$^2$,
    V~Tikhomirov$^8$,
    E~Vallazza$^4$
}

\address{$^1$Dipartimento di Fisica e Scienze della Terra, Università degli Studi di Ferrara, Ferrara, Italy}
\address{$^2$Istituto Nazionale di Fisica Nucleare, Sezione di Ferrara, Ferrara, Italy}
\address{$^3$Dipartimento di Scienza e Alta Tecnologia, Università degli Studi dell'Insubria, Como, Italy}
\address{$^4$Istituto Nazionale di Fisica Nucleare, Sezione di Milano Bicocca, Milano, Italy}
\address{$^5$Dipartimento di Fisica G. Occhialini, Università degli Studi di Milano Bicocca, Milano, Italy}
\address{$^6$Dipartimento di Fisica e Astronomia, Università degli Studi di Padova, Padova, Italy}
\address{$^7$Istituto Nazionale di Fisica Nucleare, Laboratori Nazionali di Legnaro, Legnaro, Italy}
\address{$^8$Institute for Nuclear Problems of Belarusian State University, Minsk, Belarus}
\address{$^9$Department of Physics, Wuhan University, Wuhan, China}
\address{$^{10}$Dipartimento di Fisica, Università degli Studi di Milano Statale, Milano, Italy}

%

\ead{mattia.soldani@unife.it}

\begin{abstract}
Scintillating homogeneous detectors represent the state of the art in electromagnetic calorimetry. Moreover, the currently neglected crystalline nature of the most common inorganic scintillators can be exploited to achieve an outstanding performance boost in terms of compactness and energy resolution. In fact, it was recently demonstrated by the AXIAL/ELIOT experiments that a strong reduction in the radiation length inside PWO, and a subsequent enhancement in the scintillation light emitted per unit thickness, are attained when the incident particle trajectory is aligned with a crystal axis within $\sim 1^\circ$. A SiPM-based system has been developed to directly probe this remarkable effect by measuring the scintillation light emitted by a PWO sample. The same concept could be applied to full-scale detectors that would feature a design significantly more compact than currently achievable and unparalleled resolution in the range of interest for present and future experiments.
\end{abstract}

\section{Axial alignment and coherent effects}
The electromagnetic interactions high-energy particles undergo when impinging on a crystal with small angle with respect to a lattice axis, i.e. a periodic string of nuclei, heavily differ from the amorphous target case. In particular, the spectra of bremsstrahlung radiation emission by electrons/positrons and pair production (PP) by photons in crystals depend on the angle with respect to the axis, and on the input energy \cite{1986_Baier, 1989_Baryshevskii}.

In case of ultrarelativistic particles (i.e. with an input energy of $\gtrsim 10~\mathrm{GeV}$), the axial electromagnetic field in the frame of the latter is affected by the Lorentz boost: the so-called strong field (SF) regime is attained for
$$
\chi = \frac{\gamma E}{E_0} > 1 \mathrm{,}
$$
where $\gamma = \varepsilon / m c^2$ is the Lorentz factor, $\varepsilon$ is the input energy, $E$ is the axis field in the laboratory frame and $E_0 \sim 1.32 \times 10^{18}~\mathrm{V/m} $ is the Schwinger QED critical field, above which nonlinear field effects occur in vacuum \cite{2006_Buchanam}. In this regime, the interaction between charged particles and the crystalline lattice results in the emission of quantum synchrotron radiation, which features a strong and peaked enhancement in the hard part of the photon spectrum with respect to the Bethe-Heitler, i.e. incoherent, case \cite{2018_Bandiera}. Moreover, high-energy photons experience the SF too: in fact, the PP rate per unit thickness is macroscopically enhanced with respect to the amorphous case, such an enhancement being proportional to the input energy \cite{1996_Moore}. Limited-strength SF-related effects persist down to $\chi \sim 0.1$; on the other hand, a saturation is expected to occur at about 100 times the SF threshold energy, i.e. in the multi-TeV range \cite{2005_Uggerhoj}, which is beyond the current experimental upper limit \cite{2021_Soldani}. The angular range of such effects depends on the continuous electromagnetic potential associated to the axis $U_0$. It has been evaluated as \cite{1986_Baier}
$$
\Theta_0 = \frac{U_0}{mc^2} \mathrm{,}
$$
although a limited enhancement in the bremsstrahlung spectrum by electrons/positrons and in the number of ${e^+e^-}$ pairs by photons is attained also out of the $\Theta_0$ threshold and up to approximately ${1^{\circ}}$ \cite{2018_Bandiera, 2021_Soldani}.

As a consequence of the aforementioned enhancement in the bremsstrahlung and PP cross sections, when on axis, a strong acceleration of the electromagnetic shower started by an input electron, positron or photon in the medium is attained; the latter corresponds to an overall reduction of the crystal effective radiation length $X_0$ \cite{2018_Bandiera, 2021_Soldani}. Moreover, the dependence of the enhancement factor on the primary particle energy counterbalances the increase in the shower depth: as a result, the dependence of the shower energy deposit peak longitudinal position on the input energy is cancelled out almost completely \cite{2019_Bandiera, 2017_Baryshevskii}. In case of scintillating crystals such as lead tungstate (PbWO$_4$; abb. PWO), the higher number of shower particles per unit thickness corresponds to an enhancement in the overall scintillation light yield.

\section{Setup to test the light yield enhancement in an oriented PWO crystal}
Recently, a test was performed at the CERN extracted beamlines with the aim of directly measuring the radiation enhancement in crystalline PWO, demonstrating a five-fold reduction of $X_0$ for $120~\mathrm{GeV}$ electrons in an oriented $[001]$ PWO sample. Here we describe the optimized experimental setup, similar to that discussed e.g. in \cite{2018_Bandiera} and \cite{2021_Bandiera_2}, and the development of a SiPM-based readout system to measure the enhancement in the scintillation light output.

PWO has a tetragonal, scheelite-like crystalline structure, with lattice constants $a = b = 5.456$~{\AA} and $c=12.020$~{\AA}. The SF is attained at a few tens of $\mathrm{GeV}$ and $\Theta_0 \sim 1~\mathrm{mrad}$ \cite{2017_Baryshevskii}. The sample under study in this work is a $4~\mathrm{mm}$ (i.e. $0.45X_0$) thick strip, with a transverse surface of $55 \times 2~\mathrm{mm}^2$.

\begin{figure}[htbp]
\begin{center}
    \includegraphics[width=\linewidth]{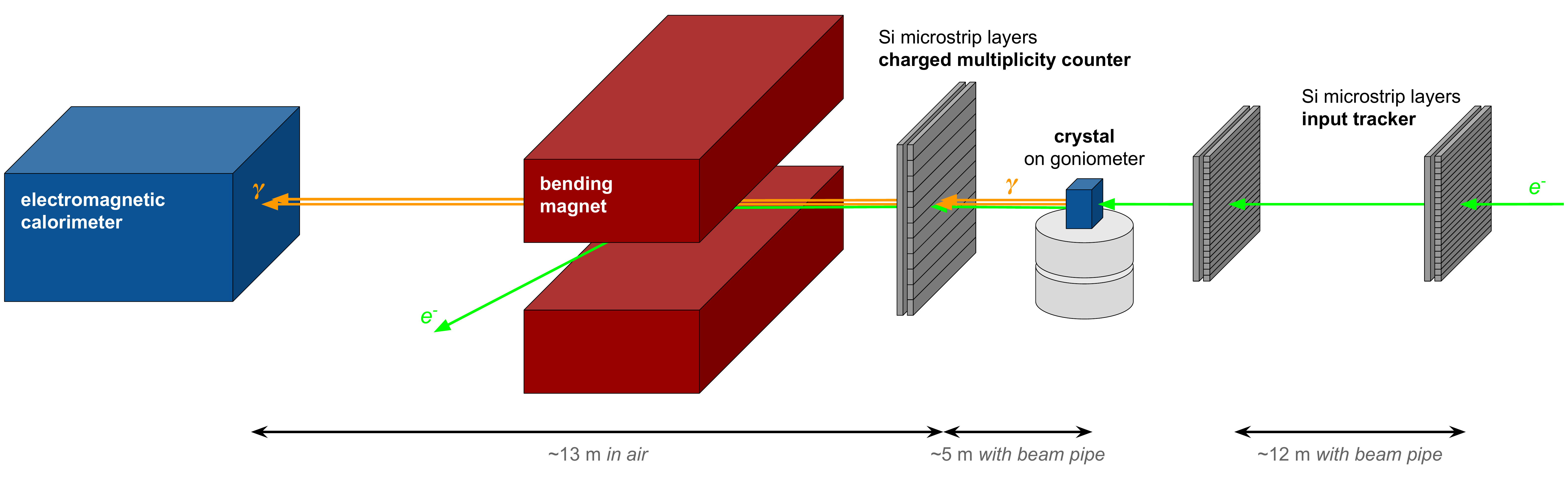}
    \caption{Experimental setup. The trajectory of the input particles is reconstructed via a silicon microstrip tracking system. Another pair of microstrip layers are placed downstream with respect to the crystal and exploited as a multiplicity counter of the output charged particles; the latter are then separated from the output photons by a dipole. Finally, a lead glass-based electromagnetic calorimeter measures the total radiated energy. Adapted from \cite{2021_Soldani}.}
    \label{fig:tb_setup}
\end{center}
\end{figure}

Figure \ref{fig:tb_setup} shows a scheme of the optimized experimental setup for the SPS extracted lines, where multi-GeV pure electron beams are available. Firstly, the input particles impinge on a telescope which consists of two tracking modules. Each module features a double-side, $\sim 2 \times 2~\mathrm{cm^2}$ silicon microstrip detector, $300~\mu\mathrm{m}$ thick, with a readout pitch of $50~\mu\mathrm{m}$; an analog readout chain makes it possible to attain a single-hit resolution of $\sim 10~\mu\mathrm{m}$ \cite{Lietti13} and, given the lever arm of $\sim 12~\mathrm{m}$, a resolution on the incident angle of $\sim 8~\mu\mathrm{rad}$. Another microstrip module, which consists of two single-side, $\sim 10 \times 10~\mathrm{cm^2}$, $410~\mu\mathrm{m}$ thick sensors, is placed downstream with respect to the target, in order to measure the multiplicity of charged particles in the output state by counting the number of hits. Finally, after the charged particles are deflected by a dipole magnet, a lead-glass calorimeter measures the energy of the photons, i.e. the energy loss in the crystalline sample.

\begin{figure}[htbp]
\centering
\begin{subfigure}{.51\textwidth}
    \centering
    \includegraphics[width=1.04\linewidth]{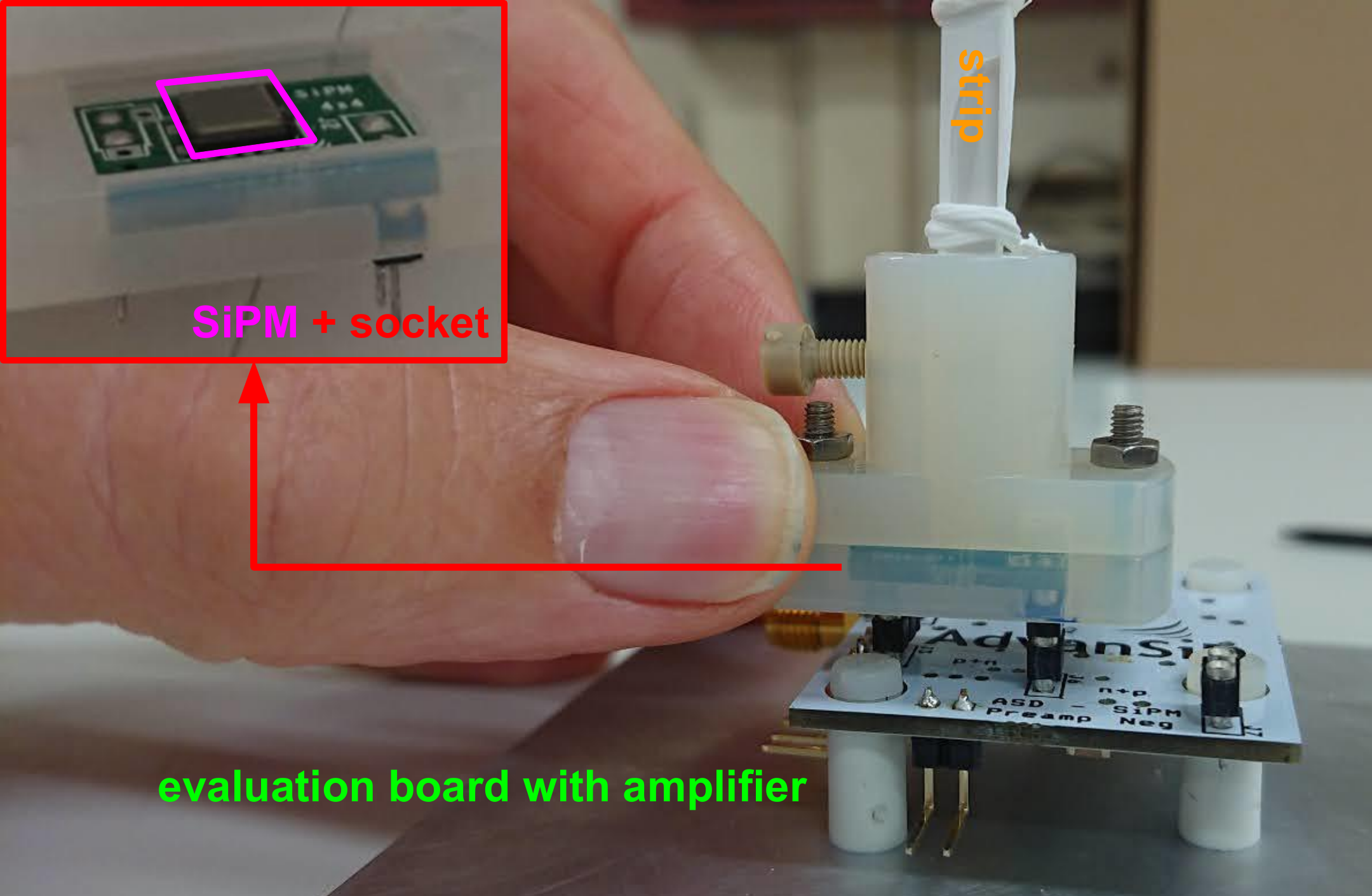}
\end{subfigure}
\begin{subfigure}{.48\textwidth}
    \centering
    \includegraphics[width=\linewidth]{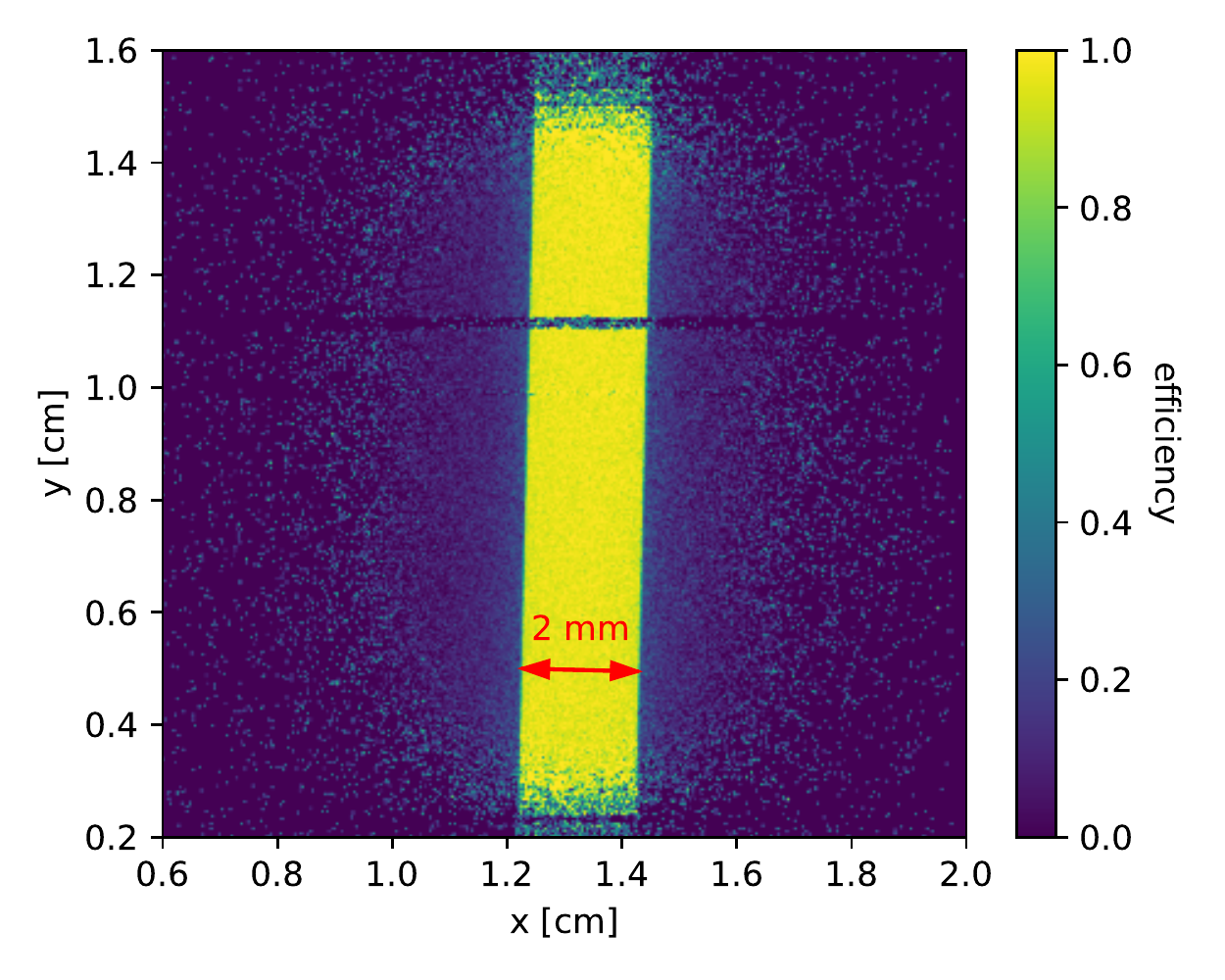}
\end{subfigure}
\caption{Left: The PWO strip (orange) coupled to the optical readout system; the detail of the SiPM (magenta) is shown in the top left corner. Right: PWO sample detection efficiency map as a function of the electron impact position.}
\label{fig:strip}
\end{figure}

A high-precision goniometer allows for the remote control of the crystal position and orientation in both the horizontal and vertical directions with a resolution of $\lesssim 5~\mu\mathrm{m}$ and $\lesssim 5~\mu\mathrm{rad}$ respectively \cite{2018_Bandiera, Lietti13, Bandiera13}.  As shown in Figure \ref{fig:strip} Left, the PWO strip is installed on the latter with the long side parallel to the vertical direction and one of the $2 \times 4~\mathrm{mm}^2$ faces, which has been polished, oriented downwards, to face the optical light readout system.

The core of the photodetection system is a Silicon PhotoMultiplier (SiPM). The {ASD-NUV4S-P} model by AdvanSiD \cite{ASD-NUV4S-P} has been chosen: its $4 \times 4~\mathrm{mm}^2$ square-shaped surface well matches the PWO strip polished face, and its photodetection efficiency, which has a FWHM of $\sim 190~\mathrm{nm}$ that ranges in {$380$--$570~\mathrm{nm}$} and the peak sensitivity wavelength at $\sim 420~\mathrm{nm}$ (NUV stands for Near UltraViolet), matches the PWO scintillation spectrum perfectly --- the latter being peaked at $420~\mathrm{nm}$ \cite{Annekov02}. Moreover, the good timing performance allows for a detailed study of the signal development.

The SiPM is coupled to the evaluation board via a dedicated socket \cite{ASD_socket}. The AdvanSiD {ASD-EP-EB-N} Evaluation Board is used in order to provide easy interface to the acquisition chain, current to voltage conversion and signal amplification \cite{ASD-EP-EB-N}. The output waveform is then acquired by a fast digitizer, the CAEN V1730C, which features a 14-bit resolution, a $500~\mathrm{MS/s}$ sampling rate and an input dynamic range that can be set either to $0.5$ or to $2~\mathrm{V}_\mathrm{pp}$ \cite{CAEN_V1730C, 18_Soldani}.

This experimental setup allows to directly measure the scintillation light signal and to study its dependence on the features of the incident particle trajectory on an event-by-event basis. It was first exploited in a beamtest in 2018 at the SPS H2 beamline. For example, Figure \ref{fig:strip} Right shows the PWO strip detection efficiency, defined as the ratio between the number of events in which the scintillation light was detected and the total number of crossing events, as a function of the hit position on the crystal surface, obtained by projecting the input electron trajectory reconstructed by the upstream tracking modules. Indeed, it is evident from Figure \ref{fig:strip} Right that an efficiency of $\gtrsim 95\%$ is attained uniformly in the whole sample under test.

\section{Conclusions and outlook}
The prototype of a custom, SiPM-based light readout system to directly measure the light emitted by oriented scintillating crystalline samples, such as PWO, and the setup adapted for the CERN SPS hundred-GeV electron beamline have been described. The outcome of this work might pave the way to the development of full-scale, ultra-compact and high-resolution homogeneous calorimeters, thanks to the strong reduction of the effective radiation length due to coherent effects in oriented crystalline scintillators. This design would prove particularly beneficial in fixed-target experiments and in space-born $\gamma$ observatories.

\section*{Acknowledgements}
We acknowledge the support of the PS/SPS physics coordinator and of the CERN SPS and EN-EA group technical staff. This work was partially supported by INFN through the PHOTAG, ELIOT and STORM experiments. 



\end{document}